\documentclass[preprint,review,12pt]{elsarticle}

\usepackage{amssymb}
\usepackage{amsmath}
\usepackage{CJKutf8}
\usepackage[overlap, CJK]{ruby}
\newenvironment{SChinese}{%
\CJKfamily{gbsn}%
\CJKtilde
\CJKnospace}{}

\newcommand{\etal}     {{\textit{et al.}\ }}

\journal{Journal of Computational Physics}

\begin{document}

\begin{frontmatter}



\title{Dispersion-dissipation condition for finite-difference schemes}


\author[label1]{X. Y. Hu} 
\author[label1]{V. K. Tritschler}
\author[label2]{S. Pirozzoli} 
\author[label1]{N. A. Adams}

\address[label1]{Lehrstuhl f\"{u}r Aerodynamik und Strömungsmechanik, Technische Universit\"{a}t
M\"{u}nchen
\\ 85748 Garching, Germany}
\address[label2]{Universit\'{a} degli Studi di Roma `La Sapienza', 
Dipartimento di Ingegneria Meccanica e Aerospaziale, Via Eudossiana 18, 00184 Roma, Italy}
\begin{abstract}
A general dispersion-dissipation condition for finite difference schemes is derived by analyzing the numerical dispersion and dissipation of explicit finite-difference schemes. The proper  dissipation required to damp spurious high-wavenumber waves in the solution is determined from a physically motivated relation between group velocity and dissipation rate. The application to a previously developed low-dissipation weighted essentially non-oscillatory scheme (WENO-CU6-M2) [X. Y. Hu and N. A. Adams; Scale separation for implicit large eddy simulation, J. Comput. Phys. 230 (2011) 7240-7249] demonstrates that this condition can serve as general guideline for optimizing the dispersion and dissipation of linear and non-linear finite-difference schemes. Moreover, the improved WENO-CU6-M2 which satisfies the dispersion-dissipation condition can be used for under-resolved simulations. We demonstrate this capability by considering transition to turbulence and self-similar energy decay of the three-dimensional Taylor-Green vortex. Simulations of the inviscid and the viscous Taylor-Green vortex at Reynolds numbers ranging from $Re=400$ to $Re=3000$ show a significant improvement over the classical dynamic Smagorinsky model and demonstrate competitiveness with state-of-the-art implicit LES models, while preserving shock-capturing properties.
\end{abstract}


\end{frontmatter}


\section{Introduction} \label{sec:intro}
The capability  of a discretization scheme for the Euler and Navier-Stokes equations to resolve small-scale wave-like solutions as well as being shock-capturing is a long-standing problem for compressible turbulent flows. Adams and Shariff \cite{AdamsShariff96} have proposed a coupling of high-resolution shock-capturing schemes with spectral-like compact finite-differences. Unlike with their original formulation \cite{Lele1992} it was necessary, however, to introduce some amount of dissipation at large wavenumbers by an upwind formulation of these compact schemes. Later, Pirozzoli \cite{pirozzoli2002conservative} pointed out that a certain amount of dissipation is desirable to damp high-wavenumber waves with incorrect propagation speed. The problem, however, is to determine how much dissipation is sufficient to eliminate spurious high-wavenumber waves, without deteriorating the physically relevant resolved wavenumber range.\par
It is a well-established fact \cite{vichnevetsky1982fourier} that the dispersive errors of finite-difference schemes are maximum near the Nyquist cut-off wavenumber, so that marginally resolved waves can spread throughout the domain and can cause non-linear instability or unphysical solutions. While current research, e.g. \cite{sun2011class}, focuses on controlling the necessary amount of dissipation, the link between dispersion and dissipation so far has not been established in terms of a quantitative criterion. \par
Since the pioneering work of Tam and Webb \cite{tam1993dispersion} on computational aero-acoustics, spectral analysis \cite{vichnevetsky1982fourier} has been widely applied to the design of high-order numerical schemes, which are especially important for numerical simulations of turbulent flows, where the accurate resolution of small scales in both amplitude and phase is 
crucial \cite{weirs1997optimization, wang2001optimized, ponziani2003development, zonglin2004dispersion, martin2006bandwidth}. A spectral analysis of finite-difference schemes shows that numerical dispersion and dissipation errors are related to the propagation of simple waves. While this spectral analysis is applicable to linear schemes, Pirozzoli \cite{pirozzoli2006spectral} developed a generalized spectral analysis for non-linear finite-difference schemes, such as shock-capturing schemes. Additionally, Sun~\etal \cite{sun2011class} recently showed that dispersion and dissipation of a class of explicit linear schemes to some extent can be controlled independently.\par
The objective of the present work is to show that the proper dissipation of a numerical scheme can be determined from a dispersion-dissipation condition. For demonstration, we apply this condition 
to diminish spurious waves produced by our previously developed WENO-CU6-M2 scheme \cite{hu2011scale}. Two straightforward modifications of the algorithm are proposed and 
verified by numerical examples.
\section{Dispersion-dissipation condition}\label{sec:method}
The fundamental concept of a dispersion-dissipation condition can be developed from the one-dimensional linear advection equation
\begin{equation}\label{advection-equation}
\frac{\partial u}{\partial t} + c\frac{\partial u}{\partial x} = 0 \quad , \quad -\infty < x < +\infty 
\end{equation}
with an initial condition consisting of a monochromatic sinusoidal wave $u(x,t=0) = e^{i\xi x}$ with wavenumber $\xi$ and unit amplitude, and $c$ as the advection velocity \cite{vichnevetsky1982fourier}. 
Eq.~(\ref{advection-equation}) is discretized in the spatial domain such that $x_j = j h$, $j=0, ..., N$, where $h$ is the constant grid spacing, 
and $u_j =u(x_j)$ is the corresponding grid function. The semi-discretized form by the method of lines yields a system of ordinary differential equations
\begin{equation}
\frac{d u_j}{d t} = - c u'_j, \quad j=0, ..., N, \quad u_j(t=0) = e^{ij \xi h} \quad , 
\label{ode}
\end{equation}
where $u'_j$ is a finite-difference approximation of the first spatial derivative at $x_j$ 
\begin{equation}
u'_j = \frac{1}{h}\sum^{r}_{l=-q} a_l u_{j+l} \quad . 
\label{ode-discretization}
\end{equation}
$a_l$ are the coefficients of the scheme that satisfy the appropriate consistency conditions. Using the linear scheme in Eq.~(\ref{ode-discretization}) the analytical solution of Eq.~(\ref{ode}) is
\begin{equation}
u_j (t) = e^{\tilde{\xi}_I c t} e^{i\xi(j h -  c t \tilde{\xi}_R / \xi )} \quad ,
\label{ode-solution}
\end{equation}
where
\begin{equation}
\tilde{\xi}(\xi) = \tilde{\xi}_R + i \tilde{\xi}_I = - \frac{i}{h} \sum^{r}_{l=-q} a_l e^{il\xi h} = \frac{1}{h} \sum^{r}_{l=-q} a_l\left[\sin(l\xi h) - i\cos(l\xi h)\right]
\label{modified-wavenumber}
\end{equation}
is the modified wavenumber. \par
Upon comparing Eq.~(\ref{ode-solution}) with the exact solution of Eq.~(\ref{advection-equation}) at $x_j$
\begin{equation}
u(x_j,t) = e^{i\xi(j h - ct)} \quad ,
\label{ode-exact}
\end{equation}
it is easy to see that the finite-difference approximation Eq.~(\ref{ode-discretization}) introduces two kinds of numerical error. The numerical dissipation $\tilde{\xi}_I$ modifies the wave amplitude by $e^{- \lambda t}$, where the decay rate $\lambda = -\tilde{\xi}_I c$ corresponds to an inverse characteristic decay time scale. Note that this time scale is sensible only for $\tilde{\xi}_I \leq 0$. The numerical dispersion $\tilde{\xi}_R / \xi$ modifies the wave phase speed by $c^*(\xi) = c~\tilde{\xi}_R / \xi$. \par
The velocity of a wave packet centered at wavenumber $\xi$ is given by the group velocity $\mathcal{V} = \mathrm{d} (\xi c^*(\xi))/\mathrm{d}\xi = c~\mathrm{d} \tilde{\xi}_R / \mathrm{d} \xi$ 
\cite{vichnevetsky1982fourier, colonius2004computational}. Spurious behavior is generated as the wave packet is transported with a velocity that differs from the exact phase velocity by $|\Delta c|= |\mathcal{V} - c|$. The central assumption for the dispersion-dissipation condition proposed in this paper is now to require that the characteristic time scale of advection of a spurious wave $r h / |\Delta c|$ is of the same order of magnitude as the characteristic decay time scale 
\begin{equation}
 r \frac{h}{|\Delta c|} \approx \frac{1}{\lambda} \quad ,
\label{dispersion-dissipation-condition1}
\end{equation}
where $r \geq 0$ is a parameter. This condition implies that the characteristic distance a spurious wave can travel away from the exactly advected wave is on the order of $r$ times the grid spacing $h$. Under this condition one derives from Eq.~(\ref{dispersion-dissipation-condition1}) that 
\begin{equation}
 0 \leq \frac{\left| \frac{\mathrm{d} \tilde{\xi}_R}{\mathrm{d} \xi} - 1 \right|}{-\tilde{\xi}_I} = r
\label{dispersion-dissipation-condition2}
\end{equation}
Eq.~(\ref{dispersion-dissipation-condition2}) is the formal expression of the dispersion-dissipation condition. If the discretization scheme is such that $r$ is large the solution can be contaminated by spurious waves that travel significant distances away from the exactly advected wave. If, on the other hand, $r$ is small the solution may be free of spurious waves, but the dissipative error is larger than necessary and can affect the resolved scales. An optimal condition determined by $r$ ensures a resaonable balance between dispersion and dissipation of spurious wave components. \par
Although the dispersion-dissipation condition has been motivated by the scalar advection equation and linear finite-difference schemes for the spatial derivatives, we show in the following section how the condition can be applied to more general, non-linear schemes, and demonstrate the feasability of the approach for increasingly complex applications.  
\section{Applications}
With the help of Eq.~(\ref{dispersion-dissipation-condition2}) it can be assessed whether a given finite-difference scheme is prone to produce spurious waves. Moreover, the coefficients of linear schemes can be adjusted such that the dispersion-dissipation condition is satisfied.
\subsection{Linear schemes}
\begin{figure}[p]
\begin{center}
\includegraphics[width=1.\textwidth]{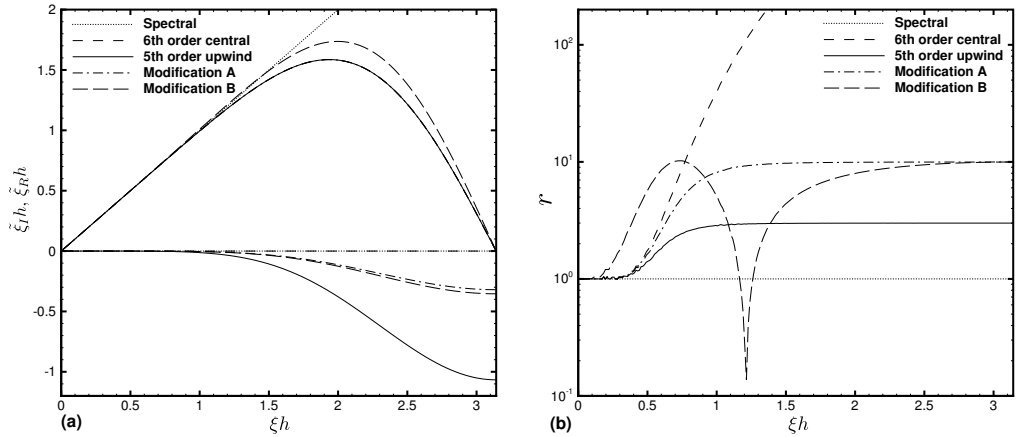}
\caption{Dispersion relation of several linear schemes: a) modified wavenumber and (b) dispersion-dissipation ratio computed with $r = \frac{\left| \frac{\mathrm{d} \tilde{\xi}_R}{\mathrm{d} \xi} - 1 \right| + \varepsilon}{-\tilde{\xi}_I + \varepsilon}$, where $\varepsilon = 10^{-3}$. } 
\label{linear-dispersion}
\end{center}
\end{figure}

\begin{figure}[p]
\begin{center}
\includegraphics[width=1.\textwidth]{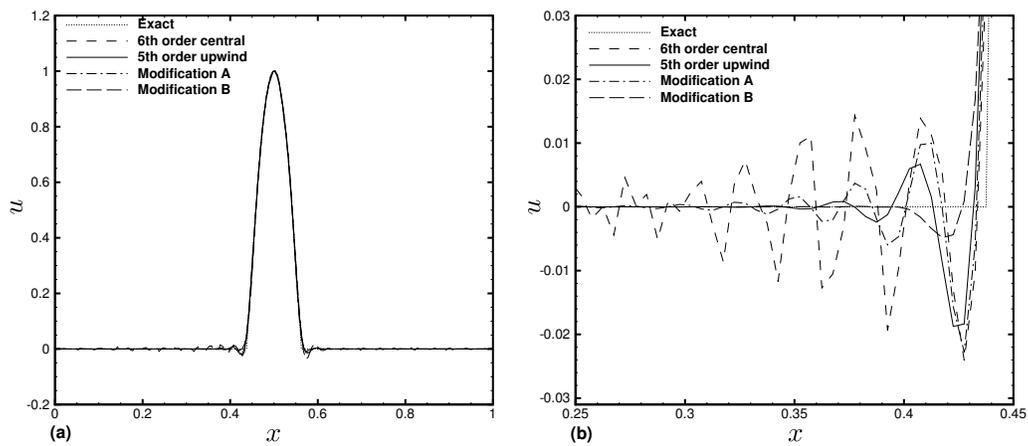}
\caption{Numerical solution of the advection of a segment of a sine wave (a) with a close-up view near the left the main front (b).} 
\label{linear-advection}
\end{center}
\end{figure}

For the explicit linear scheme in Eq.~(\ref{ode-discretization}) the condition Eq.~(\ref{dispersion-dissipation-condition2}) gives  
\begin{equation}
r = \frac{\left|\sum^{r}_{l=-q} a_l l \cos(l \xi h) -1\right| + \varepsilon}{\sum^{r}_{l=-q} a_l\cos(l\xi h) + \varepsilon} \quad ,
\label{dispersion-dissipation-condition-linear-scheme}
\end{equation}
where $\varepsilon = 10^{-3}$ to prevent division by zero. Fig.~\ref{linear-dispersion} shows the spectral properties of two typical linear schemes, a 6th-order central scheme on a 7-point stencil with $a_l = \{-\frac{1}{60},\frac{1}{20},-\frac{9}{12}, 0$, $\frac{9}{12}, -\frac{1}{20}, \frac{1}{60}\}$ and a 5th-order upwind scheme on a 6-point stencil  with $a_l = \{-\frac{1}{30},\frac{1}{4},-1$, $\frac{1}{3}, \frac{1}{2}, -\frac{1}{20}\}$.
They have the same numerical dispersion, while the central scheme is non-dissipative and the upwind scheme has substantial numerical dissipation.
As shown in Fig.~\ref{linear-dispersion}(b), the dispersion-dissipation ratio of the central scheme becomes very large for $\xi h \approx 0.8$, which suggests that spurious waves beyond this normalized wavenumber can contaminate the solution. On the other hand, the maximum dispersion-dissipation ratio of the upwind scheme is $r = 3$ which indicates that the numerical dissipation is  sufficient to suppress spurious waves. These properties can be observed from the results of the advection of a segment of a sine wave, see Fig.~\ref{linear-advection}.
Note that the spurious waves generated by the 6th-order central scheme spread throughout the entire domain, whereas for the 5th-order upwind scheme spurious waves are rapidly damped. \par
For a given stencil spurious waves can be reduced either by increasing the dissipation and leaving the dispersion unchanged (modification $A$), 
or by modifying both dissipation and dispersion (modification $B$) in order to satisfy the dispersion-dissipation relation in Eq.~(\ref{dispersion-dissipation-condition2}). For 
a 7-point stencil increasing the dissipation and leaving the dispersion unchanged results in a 5th-order linear scheme by enforcing the constraint 
$r = 10$ at $\xi h = \pi$ according to Eq.~(\ref{dispersion-dissipation-condition-linear-scheme}). The resulting coefficients of modification A are 
$a_l = \{-\frac{13}{600},\frac{9}{50},-\frac{33}{40}, \frac{1}{10}, \frac{27}{40}, -\frac{3}{25}, \frac{7}{600}\}$. 
As shown in Fig.~\ref{linear-dispersion}(a) and (b), such a modification does not change the dispersion, but achieves $r\leq10$ for the entire wavenumber range. For modification B, a 4th-order scheme is considered as the underlying scheme. The dispersion of this scheme is improved by minimizing $\int^{\xi h}_{0}e^{6(\pi - \phi)}|c^* - c|^2 d(\xi h)$ under the constraint $|c^* - c| \leq 0.015$ for all $\xi < \xi_c$, where $\xi_c = 1.5/h$ is a critical wavenumber. Dissipation of this scheme is separately constrained by enforcing $r = 10$ at $\xi h = \pi$, following  \cite{sun2011class}. The resulting coefficients of modification B are $a_l = \{- 0.032803, 0.22561, - 0.88598$, $0.11061, 0.72007, - 0.15924, 0.021742\}$.
As shown in Fig.~\ref{linear-dispersion}, the modified scheme achieves both considerable improvement with respect to dispersion, and satisfies $r\leq10$ for the entire wavenumber range.
With the above modifications the spurious waves of the original 6th-order scheme can be eliminated within a short distance from the main wavefront, see Fig.~\ref{linear-advection}.
\subsection{Non-linear schemes}
For non-linear schemes the modified wavenumber Eq.~(\ref{modified-wavenumber}) can be estimated numerically through the approximate dispersion relation (ADR) 
of Pirozzoli~\cite{pirozzoli2006spectral}. ADR still allows the dispersion-dissipation condition to be defined by Eq.~(\ref{dispersion-dissipation-condition2}).   
As an example, we consider the WENO-CU6-M2 scheme \cite{hu2011scale}, which is an adaptive central-upwind WENO scheme with 6th-order accuracy for smooth solutions. By a non-linear adaptive weighting procedure the scheme recovers either the optimal 6th-order central scheme in smooth flow regions, or a 3rd-order upwind approximation when  discontinuities are detected by the smoothness measure. The non-linear weights are given by
\begin{equation}
\omega_{r} = \frac{\alpha_{r}}{\sum^{3}_{r=0}\alpha_{r}},
\quad \alpha_{r} = d_r\left(C_q + \frac{\tau_6}{\beta_{3,r} + \varepsilon \Delta x^2}
\frac{\beta_{3,ave} + \chi \Delta x^2}{\beta_{3,r} + \chi \Delta x^2}\right)^{q},
\label{weight-6}
\end{equation}
where $d_r = \{0.05, 0.45, 0.45, 0.05\}$ are the optimal weights yielding a 6th-order linear central scheme. $q = 4$ is an integer parameter,
$C_q = 10^3$ is a positive constant parameter and $\varepsilon =1/\chi = 10^{-8}$. For the WENO methodology and the details of computing $\beta_{3,r}$, $\beta_{3,ave}$ and $\tau_6$, 
the reader is referred to~\cite{jiang1996cient, hu2011scale}. \par
\begin{figure}[p]
\begin{center}
\includegraphics[width=1.\textwidth]{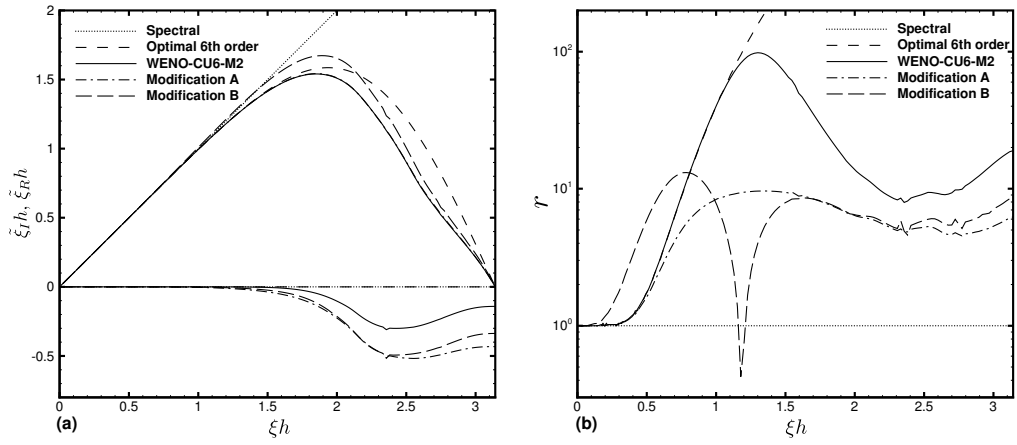}
\caption{Approximate dispersion relation of various numerical schemes:
(a) modified wavenumber and (b) dispersion-dissipation ratio computed with $r = \frac{\left| \frac{\mathrm{d} \tilde{\xi}_R}{\mathrm{d} \xi} - 1 \right| + \varepsilon}{-\tilde{\xi}_I + \varepsilon} $, where $\varepsilon = 10^{-3}$.} 
\label{dispersion}
\end{center}
\end{figure}

As shown in Fig.~\ref{dispersion}(a), the modified wavenumber of WENO-CU6-M2 is very close to that of the 6th-order linear central scheme up to $\xi h \approx \pi/2$. 
Since WENO-CU6-M2 produces dispersive errors for $\xi h \gtrsim 1$ but only weak dissipation, the dispersion-dissipation condition in Eq.~(\ref{dispersion-dissipation-condition2}) is not satisfied in the wavenumber range from $1.1 \lesssim \xi h \le \pi/2$, and reaches its maximum value of $r \approx 100$ at $\xi h \approx 1.302$, as shown in Fig.~\ref{dispersion}(b). 
Therefore, WENO-CU6-M2 is prone to produce spurious waves in the considered wavenumber range. \par
In analogy with linear schemes, WENO-CU6-M2 can be improved by changing the dispersion-dissipation property of the underlying optimal linear scheme. Instead of the 6th-order central scheme an optimized 5th-order linear scheme can be constructed by enforcing the constraint $r = 10$ at $\xi h = 1.302$ according to Eq.~(\ref{dispersion-dissipation-condition-linear-scheme}).
The resulting optimal weights of modification A are $d_r = \{0.065, 0. 495, 0.405, 0.035\}$. As shown in Fig.~\ref{dispersion}(a) and (b), such a modification does not change the dispersion,
but achieves $r<10$ throughout the entire wavenumber range. Alternatively, dispersion-dissipation optimized WENO-CU6-M2 weights can be obtained from a 4th-order linear scheme 
\cite{sun2011class} constrained by $r = 10$ at $\xi h = \pi/2$, resulting in modification B with $d_r = \{0.09045, 0.4441$, $0.39227, 0.07318\}$.
As shown in Fig.~\ref{dispersion}, the modified scheme achieves both considerable improvement in dispersion, and satisfies $r < 10$ throughout the entire wavenumber range. \par
\subsection{Validation of the optimized WENO-CU6-M2 scheme}
First, two simple numerical examples are given to illustrate that spurious waves can be produced by WENO-CU6-M2, and that such waves can be reduced by the proposed modifications. 
One example is the Sod shock tube problem, the other one is the two-blast-wave interaction problem which is taken from Woodward and Colella \cite{colella1984piecewise}.
The reference solutions for these problems are obtained on a high-resolution grid with $3200$ points using the WENO-5 scheme \cite{jiang1996cient}. All the computations are run at a CFL number of 0.6. \par
\begin{figure}[p]
\begin{center}
\includegraphics[width=1.\textwidth]{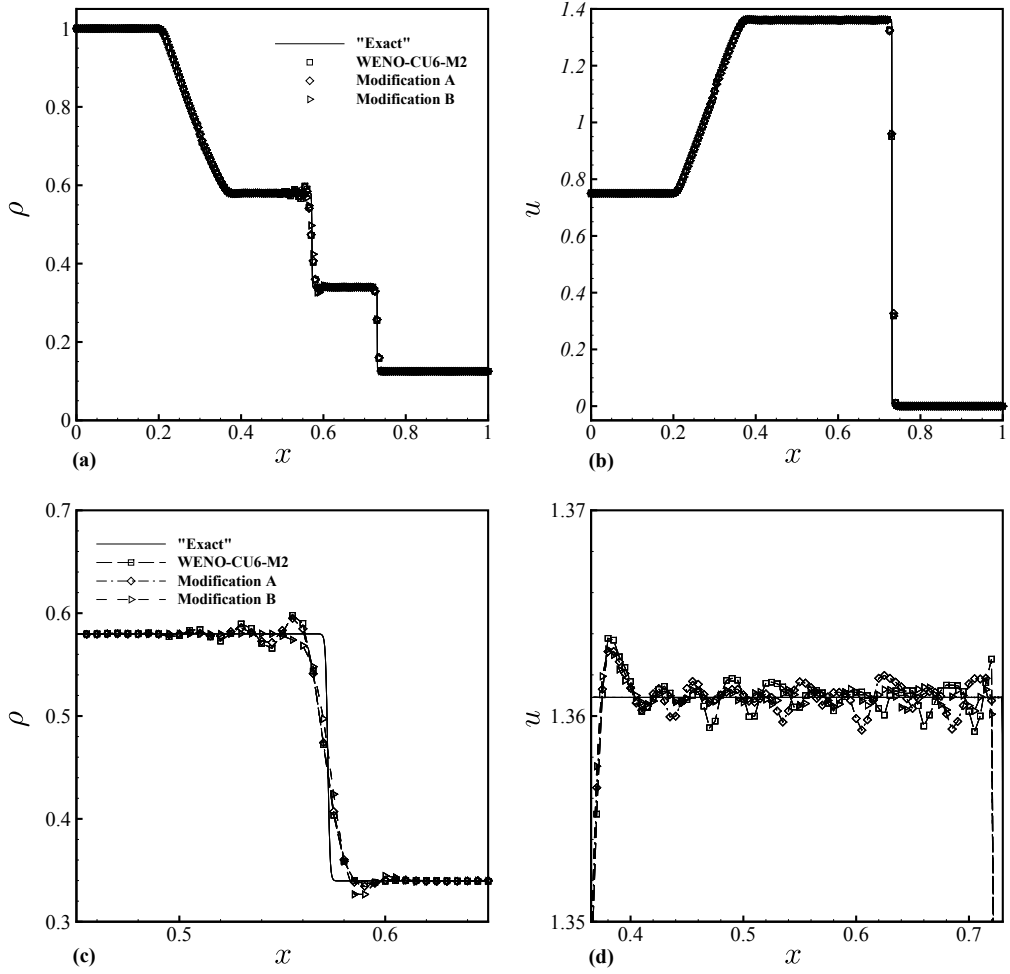}
\caption{Numerical solution of Sod's shock-tube problem at $t = 0.2$ on a grid with 200 points: (a) density and (b) velocity profiles and close-up views of 
the density (c) and velocity profiles (d).} 
\label{riemann}
\end{center}
\end{figure}

The initial condition of Sod's shock tube problem is
\begin{equation*}
\left( \rho, u, p \right) = \left\{
  \begin{array}{l l}
    \left( 1, 0.75, 1 \right) & \quad \mbox{if } 0 < x < 0.3 , \\
    \left( 0.125, 0, 0.1 \right) & \quad \mbox{if } 0.3 < x < 1.0 . 
  \end{array} \right.
\end{equation*}
The results of the original WENO-CU6-M2 scheme as well as its modifications A and B are computed on grids with 200 points using a local Lax-Friedrichs flux. While the WENO-CU6-M2 scheme 
shows good shock-capturing properties, see Fig.~\ref{riemann}(a) and (b), it also produces small oscillations at the contact discontinuity at $x \approx 0.57$, as can be seen from 
Fig.~\ref{riemann}(c). The modified versions of WENO-CU6-M2, however, damp these oscillations, while representing the contact discontinuity as sharp as the original scheme. 
The velocity profiles of modifications A and B in Fig.~\ref{riemann}(b) and (d), also show reduced spurious oscillations. \par
\begin{figure}[p]
\begin{center}
\includegraphics[width=1.\textwidth]{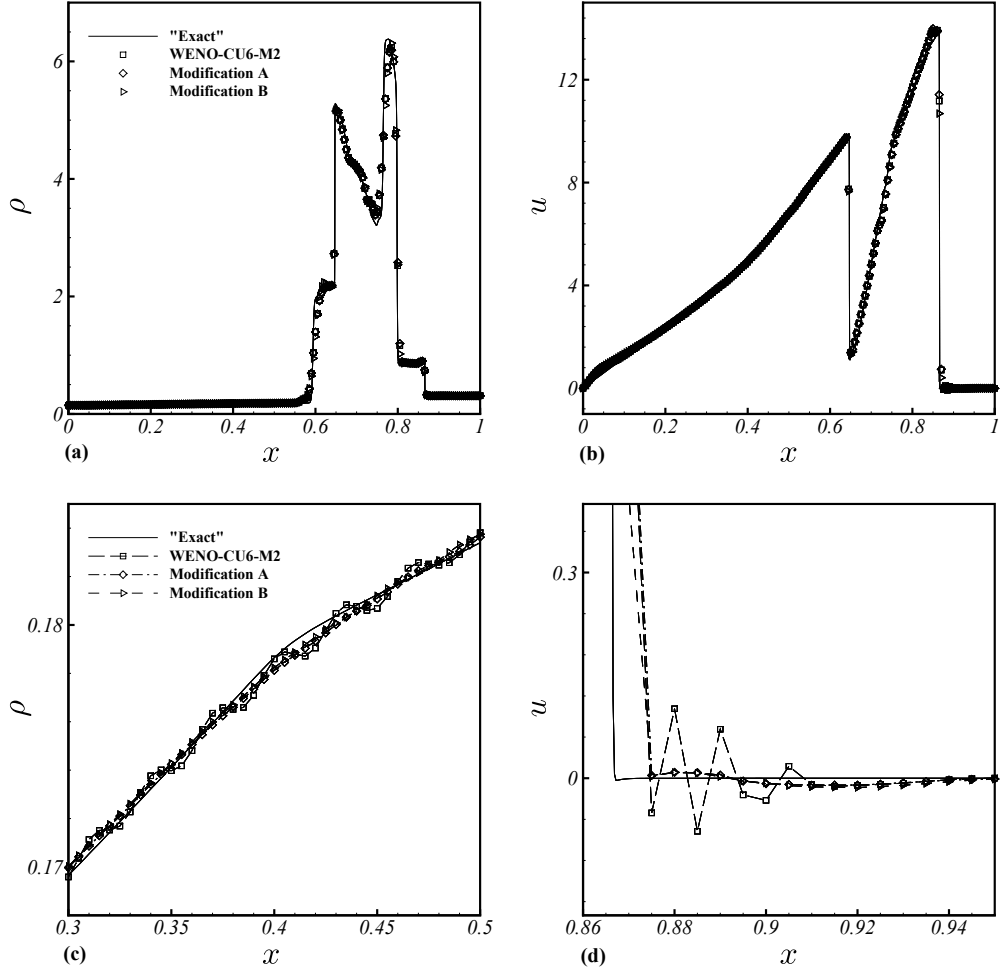}
\caption{Numerical solution of the two-blast-wave interaction problem at $t = 0.038$ on a grid with 400 points:
(a) density and (b) velocity profiles and close-up views of the density (c) and velocity profiles (d). For clarity only every second grid point is shown.} 
\label{blast}
\end{center}
\end{figure}

 
The initial condition of the two-blast-wave interaction problem is given by
\begin{equation*}
\left( \rho, u, p \right) = \left\{
  \begin{array}{l l}
    \left( 1, 0, 1000 \right) & \quad \mbox{if } 0 < x < 0.1 , \\
    \left( 1, 0, 0.01 \right) & \quad \mbox{if } 0.1 < x < 0.9 , \\
    \left( 1, 0, 100 \right)  & \quad \mbox{if } 0.9 < x < 1.0 .
\end{array} \right.
\end{equation*} 
The numerical results computed on a grid with 400 points show at first sight no differences between original and modified schemes, see Fig.~\ref{blast}. However, small spurious waves can be observed in the close-up views of the results obtained from the original WENO-CU6-M2 scheme for the density profile and for the shock wave around $x \approx 0.865$, Fig.~\ref{blast}(c) and (d). These spurious oscillations are eliminated by the two modified schemes. \par
Finally, we consider the three-dimensional inviscid Taylor-Green vortex (TGV) to examine the capabilities of the modified schemes to predict transition to turbulence. For this purpose, 
modification B is chosen as underlying linear scheme with the parameters $C$ and $q$ of the WENO-CU6-M2 scheme set to $C = 10^3$ and $q = 2$. \par
The initial condition of the TGV is given by 
\begin{align*}
 \rho &= 1 \quad , \\
 \begin{pmatrix}
  u \\
  v \\
  w 
 \end{pmatrix}
 &= 
\begin{pmatrix}
 0 \\
 \rho \cos(x) \sin(y) \cos(z) \\
 \rho \cos(x) \cos(y) \sin(z)
\end{pmatrix} \quad , \\
  p &= \frac{1}{\gamma \mbox{Ma}^2} + \frac{\rho}{16} \left[ \left( \cos(2x) + 2 \right) \left( \cos(2y) + \cos(2z) - 2 \right) \right] \quad ,
\end{align*}
with $\gamma = 1.4$ and $\mbox{Ma} = 0.1$. The TGV is solved in a cubic computational domain with side length of $2 \pi$ and discretized with $64^3$ cells. Periodic boundary conditions are applied at all sides. \par
\begin{figure}[p]
\begin{center}
\includegraphics[width=1.\textwidth]{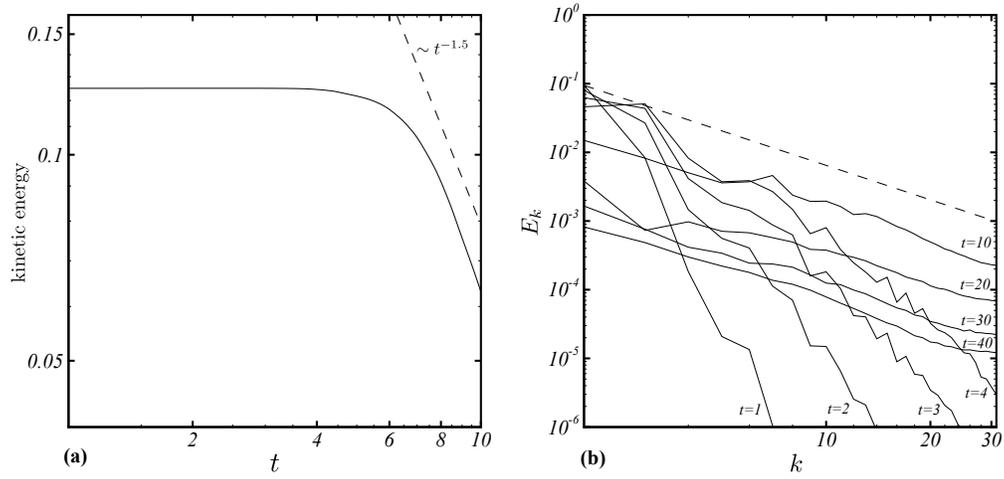}
\caption{Temporal evolution of the kinetic energy of the three-dimensional inviscid Taylor-Green vortex: (a) total kinetic energy for $t \le 10$, (b) spectral representation of the kinetic energy for times 
up to $t=40$. } 
\label{TGV_inviscid}
\end{center}
\end{figure}

At about $t \approx 4$ subgrid scales are produced resulting in a decay of kinetic energy due to subgrid-scale dissipation. The kinetic energy decay of the present 
scheme scales approximately with $\sim t^{-1.5}$ which is similar to the power-law of kinetic energy decay of the original WENO-CU6-M2 \cite{hu2011scale}, see Fig.~\ref{TGV_inviscid}(a). This decay is 
slightly steeper than the value $\sim t^{-1.2}$ found by Lesieur \cite{Lesieur2000}. Other high-resolution implicit large-eddy simulations \cite{grinstein2006recent} show a similar trend for the decay of kinetic energy. 
Between $t = 4$ and $t = 10$ turbulence develops, and at $t=10$ a well-developed Kolmogorov inertial range with $k^{-5/3}$ has established. It is worthwhile to note that the physically relevant resolved scales are unaffected by numerical dissipation and that the Kolmogorov inertial range is supported up to the cut-off wavenumber, see Fig.~\ref{TGV_inviscid}(b). During the consecutive times $t=20$, $t=30$ and $t=40$ a self-similar decay of the spectra is observed, where the Kolmogorov spectrum is further sustained and largely unaffected by the numerical scheme, see Fig.~\ref{TGV_inviscid}(b). \par
Under-resolved numerical simulations of the three-dimensional viscous TGV at Reynolds numbers $Re=400$, $Re=800$, $Re=1600$ and $Re=3000$ have been conducted on grids with $64^3$ cells. The dissipation rate of the WENO-CU6-M2, modification B, is compared to established LES models such as the dynamic Smagorinsky model and the adaptive local deconvolution model (ALDM) of Hickel~\etal\cite{Hickel2006adaptive}. As reference solution the DNS data of Brachet~\etal \cite{Brachet1984} are used. \par
\begin{figure}[p]
\begin{center}
\includegraphics[width=1.\textwidth]{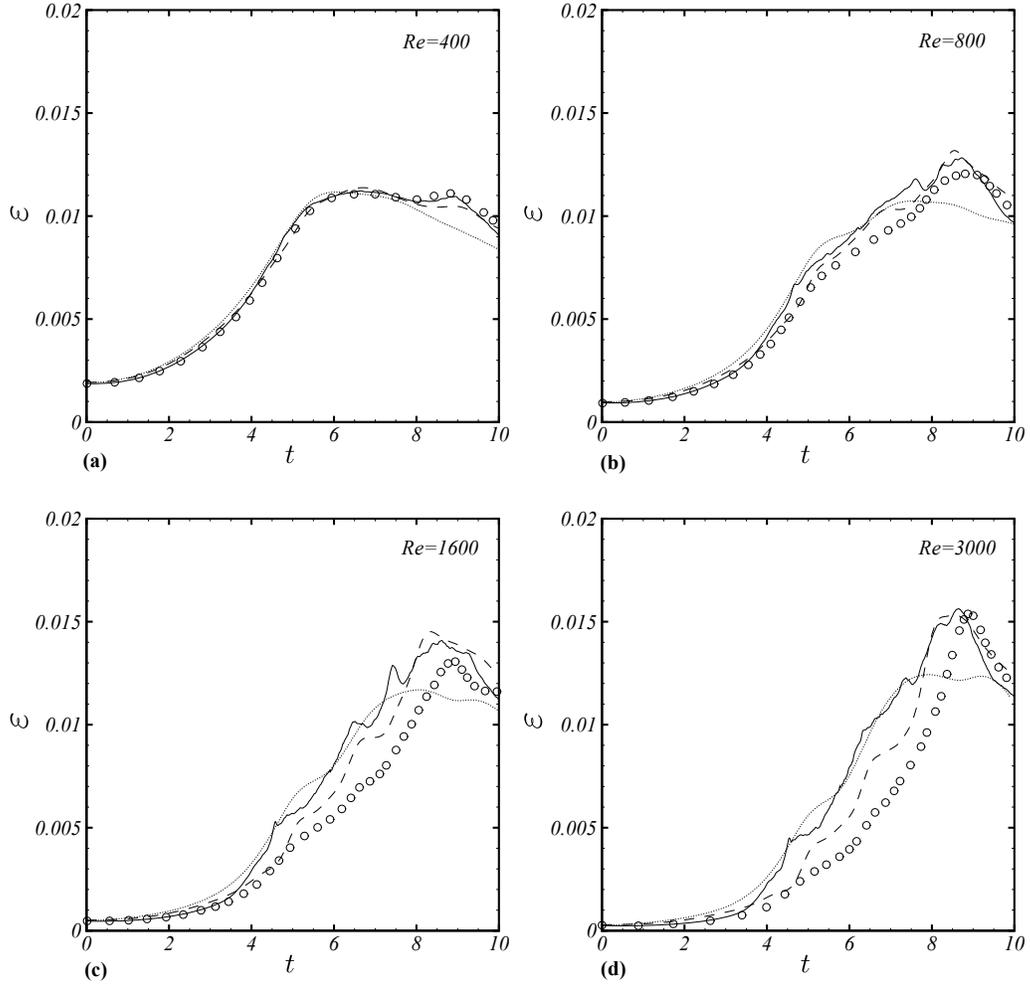}
\caption{Dissipation rates of the three-dimensional viscous Taylor-Green vortex at Reynolds numbers (a) $Re=400$, (b) $Re=800$, (c) $Re=1600$ and (d) $Re=3000$. The reference data is taken from 
Brachet~\etal (open circles). The present results (solid line), the results of the dynamic Smagorinsky LES model (dotted line) and the results of the ALDM implicit LES model (dashed line) are 
computed on $64^3$ cells. } 
\label{TGV}
\end{center}
\end{figure}

The Reynolds number specific dissipation rates $\varepsilon(t) = dE / dt$ in Fig.~\ref{TGV} indicate that the current results provide a significant improvement over the classical dynamic Smagorinsky model. Compared to ALDM it gives comparably good results for the two smaller Reynolds numbers $Re=400$ and $Re=800$. At these Reynolds numbers both models reproduce the dissipation rate of Brachet~\etal \cite{Brachet1984} At higher Reynolds numbers and early times the dissipation rates obtained with the current scheme are in excellent agreement with Brachet~\etal, while it tends to overpredict dissipation at intermediate times. The peak value of the dissipation rate is in good agreement with ALDM.
\section{Concluding remarks}
We have derived a dispersion-dissipation condition which determines the minimum numerical dissipation that a finite-difference scheme should possess to
damp spurious high-wavenumber waves in the solution. With the help of the approximate dispersion relation, this condition can also be applied to non-linear schemes. Examples 
for linear schemes and the WENO-CU6-M2 scheme suggest that this condition can be used as guidline for optimizing the dispersion and dissipation of more general grid-based linear 
and non-linear discretization schemes. \par
Furthermore, we have shown that an accordingly modified version of the WENO-CU6-M2 scheme can be used for under-resolved simulations of simple turbulent flows, as it predicts accurately the self-similar energy decay of the inviscid Taylor-Green vortex. The comparison with direct numerical simulation data of the three-dimensional viscous Taylor-Green vortex demonstrates significant improvement over the classical dynamic Smagorinsky model.
\bibliographystyle{plain}
\bibliography{turbulence}
\end{document}